\documentclass [twocolumn,prl,a4paper,altaffilletter,superscriptaddress,amsmath,amssymb]{revtex4}
\usepackage{times}
\usepackage{graphicx}
\usepackage{bm}
\begin {document}
\title{Suppression of vortex channeling in meandered YBa$_{2}$Cu$_{3}$O$_{7-\delta}$ grain boundaries}
\author {J. H. Durrell}
\email{jhd25@cam.ac.uk}
\affiliation {Department of Materials Science and Metallurgy, University of Cambridge, Pembroke
Street, Cambridge, CB2 3QZ, UK.}
\author {D. M. Feldmann}
\affiliation {Los Alamos National Laboratory, Los Alamos, NM. 87545}
\author {C. Cantoni.}
\affiliation {Oak Ridge National Laboratory, Oak Ridge, TN. 37831}

\begin {abstract} We report on the in-plane magnetic field ($H$) dependence of the critical
current density ($J_c$) in meandered and planar single grain boundaries (GBs) isolated in YBa$_{2}$Cu$_{3}$O$_{7-\delta}$ (YBCO) coated conductors. The $J_{c}(H)$
properties of the planar GB are consistent with those previously seen
in single GBs of YBCO films grown on  SrTiO$_{3}$ bi-crystals. In the straight boundary a characteristic flux channeling regime when $H$ is oriented near the GB plane, associated with a reduced $J_c$, is seen. The meandered GB does not show vortex channeling since it is not possible for a sufficient length of vortex line to lie within it.
\end {abstract}
\maketitle
The recent development of practical high temperature superconducting wires has relied on the growth of YBa$_{2}$Cu$_{3}$O$_{7-\delta}$ (YBCO) on long
lengths of highly textured flexible metal tapes \cite{rup03,selv05}. Bi-axial texture is required to maximize the grain-to-grain alignment and thereby maximize critical current ($J_{c}$), since the $J_{c}$ across a YBCO grain boundary (GB) has been found to fall off exponentially with increasing GB angle \cite{dimo88, feld01, vere01}.  The GBs in such conductors are generally low-angle GBs where the GB misorientation angle is less than 10$^\circ$.
	
Unlike high-angle GBs the superconducting order parameter is continuous across low-angle boundaries \cite{redw99} which have been found to act as flux channels \cite{diaz98, durrgb03, gure02}.  The reduced critical current associated with low-angle GBs stems from weaker vortex pinning. Consequently, for fields rotated in the plane of the conductor,
there exists a wide range of field orientations for which the GB does not depress the observed $J_{c}$. When the field is oriented far from parallel to the GB plane, even those vortices which cross the GB are effectively pinned by large lengths of vortex within the more strongly pinned intra-grain region \cite{durrgb03,pardo07}.
	
YBCO deposited by \textit{ex situ} methods, such as the physical vapor deposition and subsequent conversion of a BaF$_{2}$ based precursor,\cite{feen91} or
the trifluoroacetate metal-organic deposition (MOD) method \cite{rup03}, have been shown capable of producing highly meandered GBs \cite{feld05,feld06}.  A meandered GB is not planar, but rather consists of a complex two-dimensional interface between grains that wanders both through the thickness of
the film and along the template GB. In contrast, when the YBCO is grown by pulsed laser deposition (PLD), the GBs which form tend to be macroscopically (on length scales $>$ 100 nm) planar \cite{feld06}.
	
Previous studies of the angular dependence of the $J_{c}(H)$ properties of single YBCO GBs have, heretofore, only been carried out on PLD-grown YBCO deposited on bi-crystal substrates. This results in planar, typically pure (001) tilt, GBs \cite{diaz98,diaz98b,durrgb03}.  In this letter we report a study of the angular dependence of $J_{c}(H)$ in single meandered and planar GBs in coated conductors.  We observe a flux channeling regime for the case of the planar GB in the PLD-grown film, associated with a suppressed $J_{c}$. The meandered GB of the MOD-grown film exhibits no significant channeling regime, although the angular dependence of the GB is qualitatively different from that of the grain.

\begin{figure}
\includegraphics{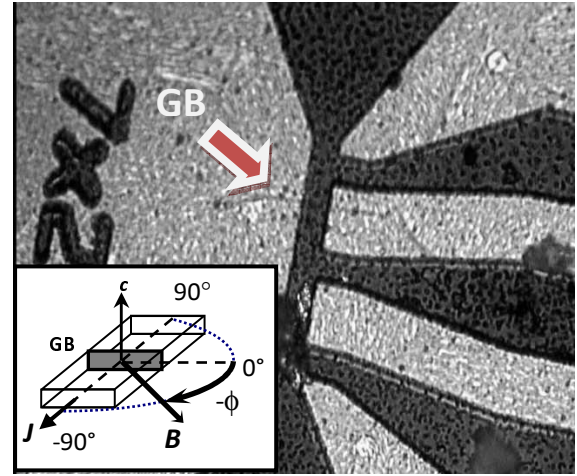}
\caption{\label{fig:1} Micrograph of a lithographically patterned
track crossing a single grain boundary, the track is 7 $\mu$m wide. The location of the grain
boundary is arrowed. The granular structure in the substrate tape
is just visible.}
\end{figure}

Single GBs were isolated and measured in YBCO films grown by PLD and MOD on rolling assisted bi-axially textured substrates (RABiTS).  Details of the procedure used to identify and isolate single grains and GBs in these RABiTS conductors, and of the RABiTS architecture, are given elsewhere \cite{feld06}. The single GBs were isolated in
tracks 7  $\mu$m wide and 20 $\mu$m long, as shown in Fig. \ref{fig:1}.  The  YBCO grown using PLD was 600 nm thick and that grown using MOD 800 nm thick.  Four-point current-voltage measurements were performed in a
two-axis goniometer mounted in a 8 T cryostat. A voltage
criterion of 750 nV was used, although different criteria did not significantly affect the angular
dependence of the critical current.  The magnetic field was rotated within the plane of the film, the in plane rotation angle $\phi$ is defined as shown in the inset to Fig. \ref{fig:1} with values of +/- 90 corresponding to the field direction aligning with the direction of macroscopic current flow.

\begin{figure}
\includegraphics{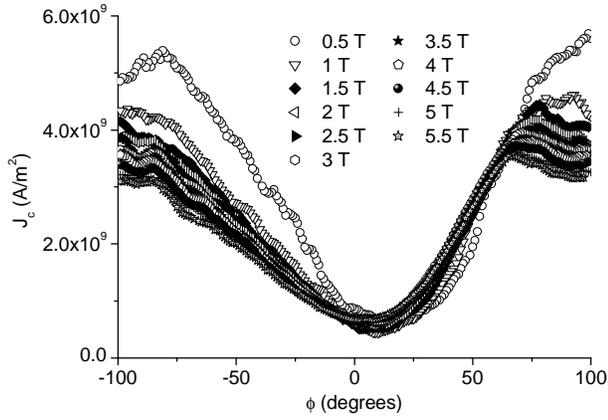}
\caption{\label{fig:2} In-plane angular dependence of the critical current in a 5$^\circ$ planar grain boundary isolated in a PLD film grown on a RABiTS substrate. On both sides of the characteristic, but more prominently for +ve $\phi$, the crossover from grain boundary dominated behavior to in-grain like behavior occurs at an angle of 65$^\circ$ from $\phi=0$ where the flux lies within the boundary.}
\end{figure}

Fig. \ref{fig:2} presents $J_{c}(\phi,H)$ for an approximately 5$^{\circ}$ planar GB in a PLD-grown YBCO film on RABiTS.  There are two regimes of behavior over the measured angular range of $\phi$.  In the central region of the plot (-65$^{\circ} < \phi < $65$^{\circ}$), the $J_{c}(\phi,H)$ curves are largely overlapping, indicating a very weak dependence of $J_{c}$ on applied field.  Beyond $|\phi| > $70$^{\circ}$ there is a much stronger dependence of $J_{c}$ with $H$. The crossover from a strong to weak field dependence of $J_{c}$ is consistent with a crossover from intra-grain dominated to GB dominated behavior, as has been previously seen in PLD-grown YBCO films on bi-crystal substrates \cite{durrgb03}. More generally suppression of field dependence has been associated with GB behavior in other measurement geometries \cite{fern03,kim05}.	

The crossover owes its existence to the previously discussed fact that low angle grain boundaries in YBCO can be simply described as a thin sheet of material where the available pinning force is low, but the order parameter is not completely suppressed.  When the flux line lies fully within the weakly pinned plane it requires only a small component of the Lorentz force to point along the weak pinning direction for the vortices to move, giving dissipation. The extended width of the channeling minimum arises from flux cutting, which permits small segments of otherwise strongly pinned vortices to cut from the vortex line and move along the channel \cite{durrgb03,pardo07}. In this case the boundary does exhibit a small amount of meandering which tends to flatten the minimum due to flux channeling.
	
\begin{figure}
\includegraphics{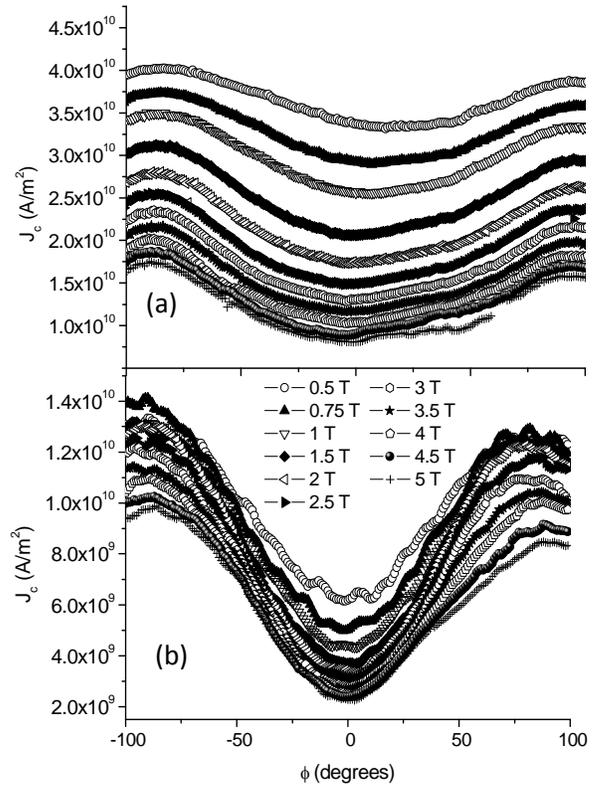}
\caption{\label{fig:3} Dependence of critical current on in-plane field orientation for (a) a track inside a single MOD template grain and (b) a 6$^\circ$ meandered boundary isolated in a MOD film. The legend applies to both parts of the figure.}
\end{figure}
	
	Figs. 3a and 3b present $J_{c}(\phi,H)$ for a single grain and single meandered GB, respectively, in an MOD-grown film on RABiTS.  The GB angle was approximately 6$^{\circ}$.  In striking contrast to the data for the planar GB shown in Fig. 2, the meandered GB (Fig. 3b) shows no evidence of a regime with a weak field dependence, associated with flux channeling.  Throughout the measured range of $\phi$ there is a significant dependence of $J_{c}$ on $H$.  However, comparing the magnitudes and shapes of the curves from the meandered GB with the associated grain measurement (Fig. 3a), it is clear there is still a suppression of $J_{c}$ due to the GB.
	
	The superior performance of the meandered grain boundary could be attributed to the increased cross sectional area of the grain boundary \cite{dinn07}. While this effect has been previously observed and is not excluded by the results presented in this letter, we do not believe an increase of area alone is sufficient to explain the present results. An increased area would increase the magnitude of $J_c$, but would not completely eliminate a region of weak dependence of $J_c$ on applied field. The absence of such a region in the data of the meandered GB indicates that the flux channeling/cutting effect is suppressed for this case.

 We propose that this is due to the interaction between the vortices and the meandered boundary. For a given transport current, the criteria controlling the stability of a vortex segment within the GB is the length of that segment. \cite{pardo07}  The shorter the vortex segment in the GB, the greater the transport current required to cut the the segment from its longer vortex line.  For a planar GB this is a simple function of the magnetic field orientation \cite{durrgb03}. For the meandered GB, the meander amplitudes have been shown to be in excess of 1 $\mu$m in films similar to the one presented here \cite{feld06}.  Therefore, at any given field orientation, only a short length of the vortex line may lie in the GB, increasing the Lorentz force and thus transport current required for vortex cutting.  This is shown in the plan-view schematics of Fig 4.  The meander also results in adjacent vortices crossing the GB at different points along their length, thus the distance between GB vortex segments in adjacent vortices through the film may now be greater than the inter-vortex spacing, further suppressing vortex cutting and channeling.

 \begin{figure}
\includegraphics{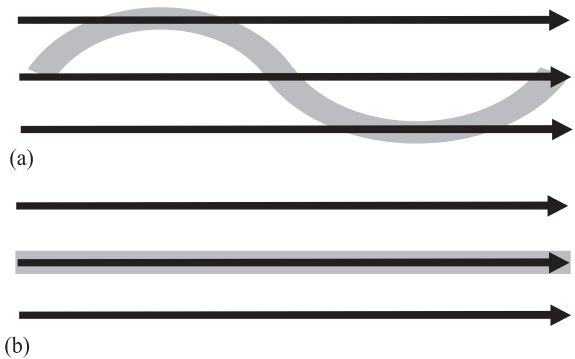}
\caption{\label{fig:4} Plan-view schematic interaction between a flux vortex and (a) meandered and (b) straight GBs. The black arrows indicate the direction of applied field, and thus vortices, the grey line represents the grain boundary. Only a fraction of a vortices can lie in the GB plane at any one time for the case of the meandered GB, consequently the maximum force per vortex segment, which is the criteria for vortex cutting, is limited \cite{pardo07}. Conversely for the straight boundary one vortex line can lie fully within the grain boundary, resulting in the entire line being subject only to the weak volume pinning force at the boundary. In previous studies \cite{feld06} it has been shown that the meander extends over a distance of up to 10 micron either side of the mean direction of the boundary.}
\end{figure}

 In conclusion, we have measured the angular dependence of the critical current in both planar and meandered GBs in YBCO films on RABiTS. The planar GB shows behavior consistent with previous studies of planar GBs on STO bi-crystal substrates, with a region of $\phi$ space associated with flux cutting/channeling and a reduced $J_c$. For the meandered GB, no fully developed flux cutting/channeling regime is achieved due to the shorter fraction of any vortex lying in the GB and geometrical complications of flux channeling due to the meander, resulting in an enhanced $J_{c}(H)$.   These results demonstrate that meandered GBs are highly beneficial for RABiTS-type coated conductors for use in high magnetic field applications.

 We thank D. C. Larbalestier and M. G. Blamire for their support of this project.  Financial support from the Royal Society, the EPSRC, the Air Force Office of Scientific Research and by the Division of Materials Sciences and Engineering, U.S. Department of Energy under contract with UT-Battelle LLC is gratefully acknowledged. The authors are grateful to AMSC for their kind provision of MOD-RABiTS samples.

\end{document}